\documentclass[sigconf]{acmart}

\usepackage[utf8]{inputenc}    
\usepackage{times}             
\usepackage{latexsym}          
\usepackage{bm}        

\usepackage{amsmath}           
\usepackage{amsthm}            
\usepackage{breqn}             

\usepackage{graphicx} 
\usepackage{booktabs}           
\usepackage{multirow}           
\usepackage{makecell}           
\usepackage{subcaption}

\usepackage{algorithm}
\usepackage{algorithmic}       

\usepackage{graphicx} 

\newcommand{\bmS}{\bm{S}}

\newcommand{\bmQ}{\bm{Q}}

\newtheorem{definition}{definition}

\newcommand{\yuya}[1]{{\color{black}#1}}
\newcommand{\kaina}[1]{{\color{black}#1}}

\AtBeginDocument{%
  }

\setcopyright{acmlicensed}
\copyrightyear{2025}
\acmYear{2025}
\acmDOI{XXXXXXX.XXXXXXX}

\acmConference[CIKM '25]{the 34th ACM International Conference on Information and Knowledge Management}{November 10--14, 2025}{Coex, Seoul, Korea}

\acmBooktitle{Proceedings of the 34th ACM International Conference on Information and Knowledge Management (CIKM ’25), November 10--14, 2025, Coex, Seoul, Korea}

\acmISBN{978-1-4503-XXXX-X/2018/06}

\begin{document}

\title{Workload acceleration by optimizing materialized view selection using local search}


\author{Kaina Anderson}
\affiliation{%
  \institution{The University of Osaka}
  \city{Suita}
  \state{Osaka}
  \country{Japan}
}
\email{anderson.kaina@ist.osaka-u.ac.jp}
\orcid{1234-5678-9012}
\author{Yohanes Yohanie Fridelin Panduman}
\affiliation{%
  \institution{The University of Osaka}
  \city{Suita}
  \state{Osaka}
  \country{Japan}
}
\email{yohanes@ist.osaka-u.ac.jp}

\author{Yuya Sasaki}
\affiliation{%
  \institution{The University of Osaka}
  \city{Suita}
  \state{Osaka}
  \country{Japan}
}
\email{sasaki@ist.osaka-u.ac.jp}

\author{Makoto Onizuka}
\affiliation{%
  \institution{The University of Osaka}
  \city{Suita}
  \state{Osaka}
  \country{Japan}
}
\email{onizuka@ist.osaka-u.ac.jp}









\begin{abstract}
The growing size of database workloads has made view selection a key performance challenge. Materializing frequent sub-queries \yuya{in workloads} improves \yuya{query efficiency, but it incurs significant view maintenance costs due to updates.}
\yuya{Although existing methods such as BIGSUBS address this trade-off between the benefit of using materialized views and the overhead of view maintenance, they have two drawbacks: insufficient maintenance cost modeling and ineffective view selection due to probabilistic techniques.}
We propose a novel view selection method that incorporates incremental view maintenance cost directly into the optimization objective \yuya{of an integer linear program} and applies local search to efficiently explore the solution space. In order to apply local search to the view selection problem, we develop neighboring solutions using sub-query containment, and select initial solutions based on sub-query frequency, utility, or utility per storage unit.
Experiments using Redbench, a benchmark simulating real-world query workloads on Amazon Redshift, show that our approach outperforms BIGSUBS in both optimization utility and the quality of selected views.
\end{abstract}




\maketitle

\section{Introduction}

In recent years, the growing number of database users has significantly increased the volume of executed queries in workloads, \yuya{making the optimization of workload performance a critical challenge}.~\cite{lan2021survey,intro:survey-answer,azure,bigsubs}. 
\yuya{A materialized view is a common database technique that stores the physically materialized result set of a query.}
Materializing the results of frequent sub-queries in workloads is a simple but effective way to significantly improve the workload performance~\cite{rela:MV_survey, halevy2001answering}. 
However, when updates are made to the underlying tables, the corresponding materialized views must be maintained, which can be time-consuming. 
\yuya{Therefore, there is a trade-off between the benefit of using materialized views and the overhead of view maintenance.}

\smallskip
\noindent
{\bf Motivation}.
Several materialized view selection methods have been proposed to balance the trade-off~\cite{bigsubs,azure,wide-deep,silva2012exploiting,zhou2007efficient}.
Specifically, BIGSUBS \cite{bigsubs} is designed for large-scale workloads, which accelerates view selection by iteratively choosing 1) which sub-queries to materialize and 2) which sub-queries to use for evaluating the queries in the workload.
In addition, BIGSUBS limits the total size of selected views by imposing a storage capacity constraint, expecting that it mitigates significant performance degradation during updates. 
However, there are two issues with BIGSUBS caused by its design, 
1) the storage capacity constraint is not sufficiently effective to reflect the overhead of view maintenance, and 2) the quality of solutions found by BIGSUBS tends to be low because it relies on probabilistic techniques for view selection.

\smallskip
\noindent
{\bf Contribution}.
To address the above two issues, we propose a novel method that balances the trade-off between the benefit of using materialized views and the overhead of view maintenance.
Our method offers two key contributions.
First, by incorporating incremental view maintenance costs into the objective function of an integer linear program, we ensure that the overhead of view maintenance is directly reflected in the optimization process.
Second, we employ a well-known optimization heuristic, local search~\cite{aarts1997local}, which effectively explores the solution space to find high-quality approximations.
In detail, applying local search to the materialized view selection poses two technical challenges: defining appropriate neighboring solutions and selecting suitable initial solutions.
To define neighboring solutions, we leverage the insight that sub-queries with a containment relationship tend to provide similar benefits of using materialized views for workloads. We treat two sub-queries are neighbors if one contains the other or is contained by the other.  
For initial solution selection, we use simple three methods based on sub-query characteristics: high frequency, high utility, and high utility per storage unit in the workload.

We conduct experiments to evaluate our proposed method in terms of optimization speed of view selection, the quality of selected views, and the workload performance using Redbench, a recent benchmark that simulates query frequencies in real-world production queries in Amazon Redshift based on the Join Ordering Benchmark (JOB) \cite{job}.
These experiments demonstrate that our approach outperforms the state-of-the-art method, BIGSUBS.

\smallskip
\noindent
{\bf Organization}.
The remainder of this paper is organized as follows. Section \ref{sec:prel} describes preliminaries. Section \ref{sec:proposal_overview} presents the overview of our workflow, and Sections~\ref{sec:proposal_viewselection} and \ref{sec:mv_create_query_rewrite} describe methods of view selection and query rewriting in detail, respectively. Section \ref{sec:exp} presents experimental results to demonstrate its effectiveness. Section \ref{sec:rela} discusses related work, and Section \ref{sec:conc} concludes the paper with a summary and directions for future work.
\section{Preliminaries} 
\label{sec:prel}




A workload consists of a set of queries, $\bmQ = \{q_1, q_2, \cdots, q_n\}$ where $q_i$ and $q_j$ for all $i,j~(i \neq j)$ are different and $|Q|=n$ denotes the number of queries in the workload.
Let $\bmS = \{s_1, s_2, \cdots, s_m\}$ be a set of sub-queries extracted from $\bmQ$ where $|S|=m$ is the number of sub-queries.
The set $\bmS$ of sub-queries is denoted as follows: 
\begin{align} 
\label{sub-queries}
\bmS = \bigcup_{q_i \in \bmQ} \texttt{extractSubQueries}(q_i),
\end{align} 
where the function $\texttt{extractSubQueries}(\cdot)$ takes a query as input and returns the set of sub-queries corresponding to partial execution plans extracted from the execution plan of $q_i$.
Since execution plans are represented as tree structures, the resulting sub-queries may exhibit containment relationships, i.e., a sub-query may include another.

\begin{figure}[h] 
    \centering 
    \includegraphics[width=\hsize]{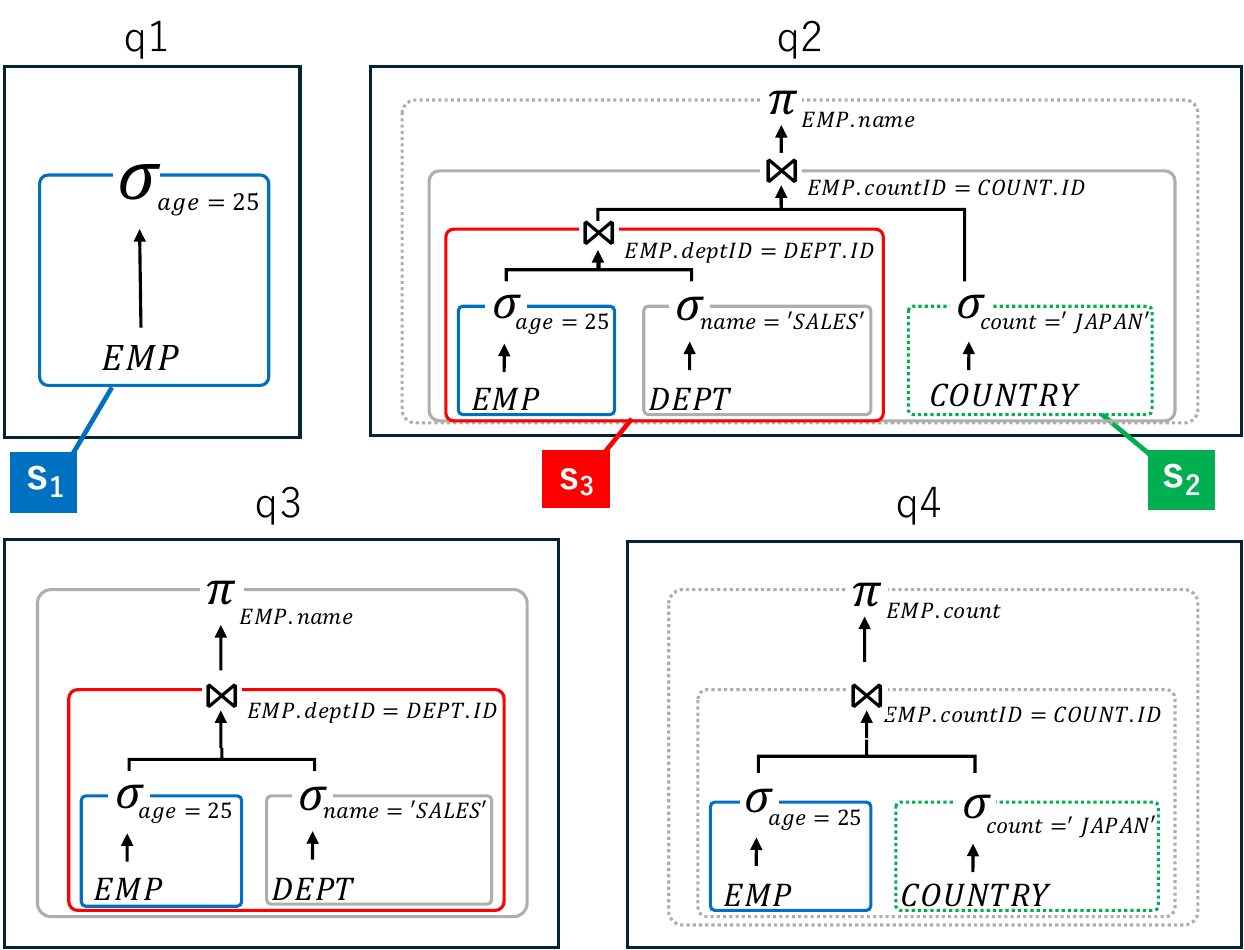} 
    \caption{An example of workload. Sub-queries shared across multiple queries: sub-query $s_1$ (blue line) is shared by four queries $q_1$-$q_4$, sub-query $s_2$ (green line) is shared by $q_2$ and $q_4$, and sub-query $s_3$ (red line) is shared by $q_2$ and $q_3$. } 
    \label{fig:subq_over_query} 
\end{figure}

\yuya{
Figure~\ref{fig:subq_over_query} shows a workload including four queries $q_1$ to $q_4$.
Each query has a tree-structured execution plan. For example, $q_2$ includes six sub-queries. Totally, there are 9 sub-queries in the workload.
}

\subsection{Sub-query Utility and Storage Size}

\yuya{
View utility~\cite{bigsubs} is a common notion that represents the effectiveness of a view for a given query and database. 
The utility of a sub-query is defined as follows:

\begin{definition}
\label{def:view_utility}
The utility of sub-query $s_j$ for query $q_i$ in a database is defined as the reduction in the estimated execution cost of $q_i$ 
using the materialized result of $s_j$:
\begin{align} \label{def:utility}
u_{ij} = \mathsf{cost_{read}}(q_i) - \mathsf{cost_{read}}(q_i|s_j),
\end{align} 
where $\mathsf{cost_{read}}(q_i)$ and $\mathsf{cost_{read}}(q_i|s_j)$ are the estimated costs of executing $q_i$ without and with utilizing the materialized result of $s_j$, respectively.
\end{definition}

The total utility $U_j$ of sub-query $s_j$ is defined as the sum of its utility for each query $q_i$ across $\bmQ$:
\begin{align} 
\label{U_j} 
U_j = \sum_{q_i\in Q} u_{ij}.
\end{align}
We hereby note that if $q_i$ does not include $s_j$, $u_{ij}$ is zero.

We define that a storage size $b_j$ of $s_j$ is the estimated size of the result executed by $s_j$.
}

\subsection{Problem Definition}

\yuya{
We solve the following problem in this paper.

\smallskip
\noindent
{\bf Problem definition}. {\it Given a database, $\bmQ = \{q_1, q_2, \cdots, q_n\}$, and storage budget $B_{max}$, we aim to find a set of sub-queries $\subseteq \bmS$ as views that minimize workload execution time and view maintenance time where the total estimated size of views is less than $B_{max}$. We assume that update operations are not given.}
\smallskip
}

Materializing common sub-queries shared across multiple queries enables improving workload performance.
As illustrated in Figure~\ref{fig:subq_over_query}, sub-query $s_1$ (blue line) is shared by four queries $q_1$-$q_4$, and sub-query $s_2$ (green line) is shared by $q_2$ and $q_4$. 
However, in addition to the benefits of reusing materialized views, we have an additional cost of view maintenance when the related base tables are updated.

This observation naturally raises a fundamental question: which sub-queries should be materialized to improve the workload performance?
While $s_1$ appears in four queries, its relatively small query plan suggests limited benefit from materialization. Conversely, $s_3$ includes an expensive join operation, which may be effective for reuse; however, it may incur higher view maintenance costs.
These trade-offs reveal the complexity of the materialized view selection problem.

\section{Overview of Our Workflow for Selecting and Using Views} \label{sec:proposal_overview}

\begin{figure}[t] \centering 
\includegraphics[width=\hsize]{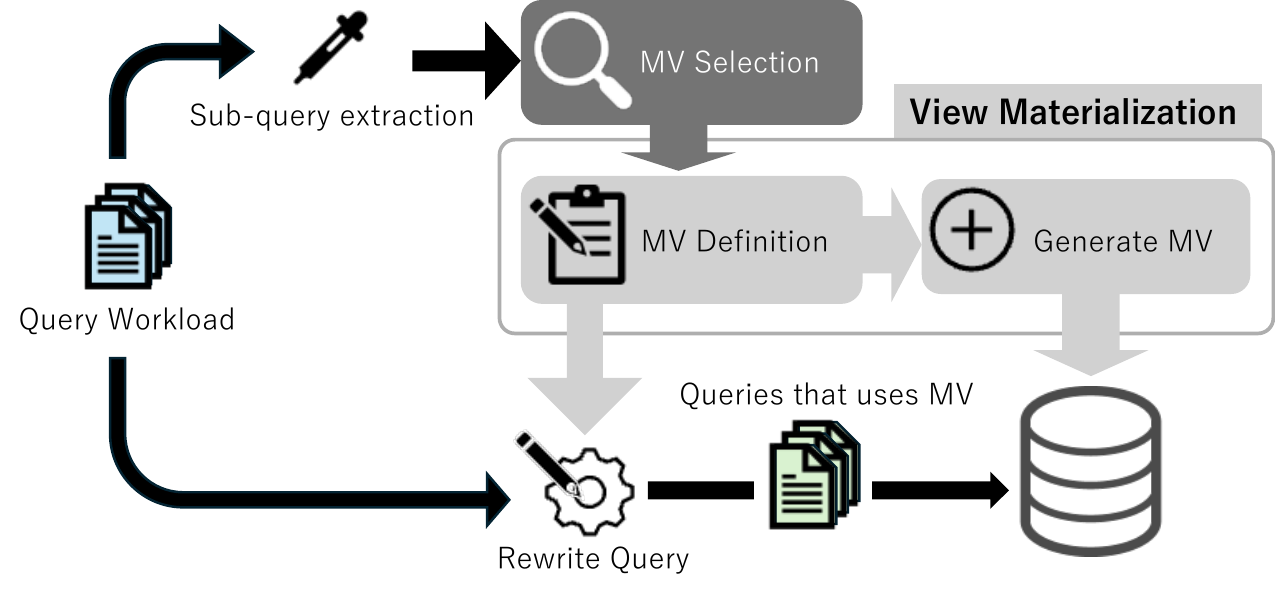} 
\caption{Workflow consisting of 
1) sub-query extraction, 2) materialized view (MV) selection, 3) view materialization, and 4) rewriting queries using materialized views.} \label{fig:proposed} \end{figure}

This section presents an overview of our workflow for selecting and using views to achieve a balance between the benefit of using materialized views and the overhead of view maintenance.
Figure~\ref{fig:proposed} provides a workflow, which consists of 
1) sub-query extraction,
2) materialized view selection, 
3) view materialization, and 
4) rewriting queries using materialized views. 

The materialized view selection offers several key contributions.
First, by incorporating incremental view maintenance costs into the objective function, we ensure that the overhead of view maintenance is directly reflected in the optimization process (Section \ref{ssec:objective}).
Second, we employ a well-known optimization heuristic, local search~\cite{aarts1997local}, which effectively explores the solution space to find high-quality approximations (Section \ref{ssec:viewSelection}).
Third, we define the incremental view maintenance cost of materialized views in response to update operations in the  workload (Section \ref{ssec:view_maintenane}), and then we describe the algorithm of the materialized view selection (Section \ref{ssec: algorithm}).
Finally, we generate materialized views and rewrite the original query to use the views (Section \ref{sec:mv_create_query_rewrite}).

\section{Method for View Selection}
\label{sec:proposal_viewselection}

We describe how to select views in detail.

\subsection{Objective Function for View Selection}
\label{ssec:objective}
The objective of materialized view selection is to balance the trade-off between the benefit of using materialized views and the overhead of view maintenance.
This problem can be formulated as an integer linear program (ILP) that maximizes the difference between the utility of sub-queries and the view maintenance cost.
\begin{definition} [View selection problem]
\label{def:loss}
Let $u_{ij}$ be the sub-query utility as defined in Equation (\ref{def:utility}), 
$m_j$ be the incremental maintenance cost of materialized sub-query $s_j$\footnote{Section \ref{ssec:view_maintenane} describes the detail of incremental view maintenance cost $m_j$.}, 
$b_j$ is the storage size of $s_j$, and $B_{max}$ be the storage budget. 
Given queries $\bmQ$ and sub-queries $\bmS$, the view selection problem is defined as follows:
\begin{align}
    L(\bmQ,\bmS)=\text{maximize } &\sum_{q_i\in \bmQ} \sum_{s_j\in \bmS} u_{ij} \cdot y_{ij} - \sum_{s_j\in \bmS} m_j \cdot z_j\notag \\
    \label{st:storage} \text{subject to } \sum_{s_j\in \bmS} b_j &\cdot z_j \le B_{max} \\
    \label{st:subq} y_{ik} + \frac{1}{|\bmS|} \sum_{j \ne k} &y_{ij} \cdot x_{jk} \le 1 \quad \forall i \in [1,|\bmQ|], \forall k \in [1,|\bmS|] \\
    \label{st:materialize} y_{ij} \le z_j \quad \forall i &\in [1,|\bmQ|], \forall j \in [1,|\bmS|]
\end{align}
\end{definition}
\noindent
$L(\bmQ,\bmS)$ is the objective function, which maximizes the sum of the utility of sub-queries over all queries (first term) minus the incremental view maintenance cost (second term)
\footnote{We refer to the contribution of each sub-query to this objective as its \textit{effective utility}.}. 
The utility of sub-queries is computed as $u_{ij} \cdot y_{ij}$, where the decision variable $y_{ij}$ indicates whether query $q_i$ uses the materialized view of $s_j$.
The view maintenance cost is computed as $m_j \cdot z_j$, where the decision variable $z_j$ indicates whether $s_j$ is selected for materialization.
Equations~(\ref{st:storage})–(\ref{st:materialize}) define the constraints of the problem:
\begin{itemize}
    \item The storage constraint (\ref{st:storage}) ensures the total size of materialized views does not exceed $B_{max}$. 
    \item The containment constraint (\ref{st:subq}) prevents overlapping views from being used in the same query. Binary variable $x_{jk}$ indicates whether $s_j$ contains $s_k$.
    \item The materialization constraint (\ref{st:materialize}) ensures that only materialized sub-queries can be used for query processing. 
\end{itemize}
\noindent
Solving this integer programming formulation yields: 1) which sub-queries $s_j$ should be materialized ($z_j = 1$), and 2) which query $q_i$ should use which sub-query $s_j$ ($y_{ij} = 1$).

\subsection{Optimization using Local Search}
\label{ssec:viewSelection}
Naively solving the optimization formulation presented in Section~\ref{ssec:objective} as an Integer Linear Programming (ILP) leads to severe performance degradation as the number of queries increases.
Indeed, ILP is known to be NP-hard and typically incurs an exponential time cost with respect to the number of decision variables, i.e., $y_{ij}$ and $z_j$ where the size depends on the product of query size  ($|\bmQ|$) and sub-query size  ($|\bmS|$).
Furthermore, although BIGSUBS~\cite{bigsubs} is designed for large-scale workloads and accelerates view selection, 
the quality of its solutions tends to be low due to its reliance on probabilistic techniques for view selection.

To address this, we employ a well-known optimization heuristic, local search~\cite{aarts1997local}, to iteratively approximate the optimal materialized views. 
Local search is a technique that avoids exhaustive solution enumeration by incrementally exploring the solution space through the enumeration of neighboring solutions to the current one, thereby enabling iterative refinement.

In order to apply the local search to the materialized view selection problem, we first generate promising initial candidate sub-queries (Section~\ref{sssec:initial}).
Then, we explore neighboring solutions by modifying the current candidate sub-queries to explore the solution space. For the expanded solutions by adding neighboring solutions, we solve an ILP to select new candidates that maximize the objective function (Section~\ref{sssec:neighbor}).
This process is repeated until no further improvement can be achieved in the objective function (Section~\ref{sssec:iteration}).


\subsubsection{Initial Solution Selection} \label{sec:initial}
\label{sssec:initial}
We propose three methods for selecting promising initial candidate sub-queries. 
We use one of the following metrics for sub-query $s_j$: 
1) frequency $F_j$ in the workload,
2) total utility $U_j$, or
3) utility per unit storage $E_j = U_j / b_j$. 
Based on the selected metric, the top $k$ sub-queries are chosen as the initial candidate sub-queries, such that the total storage cost does not exceed the storage budget $B_{\text{max}}$.

\subparagraph{\textbf{topk-}$F$:}
This method simply prioritizes sub-queries that appear most frequently in the workload. By favoring commonly shared sub-queries, it may enhance the effectiveness of selecting initial views. 

\subparagraph{\textbf{topk-}$U$:}
This method selects sub-queries with the highest total utility, as defined in Equation~(\ref{U_j}). It is expected to be more effective than topk-$F$ because it explicitly accounts for the actual benefit of utilizing materialized views. While this approach successfully identifies high-utility candidates, it may also select sub-queries that require substantial storage, potentially reducing storage efficiency.

\subparagraph{\textbf{topk-}$E$:}
This method balances utility and storage requirements by selecting sub-queries based on their utility per unit of storage.

\subsubsection{Neighborhood-based sub-query exploration}
\label{sssec:neighbor}
To explore the solution space via local search, we first define the neighbors of sub-queries.
Our approach leverages containment relationships in query plan trees: if the plan tree of one sub-query contains that of another, these two sub-queries are considered neighbors.
A parent sub-query expresses a larger plan tree that fully contains those of its child sub-queries, thereby it materializes a greater portion of the overall query processing.
In contrast, child sub-queries are more likely to be shared across multiple queries due to their smaller structure of plan trees.

We define the neighbors of sub-queries as follows.
\begin{definition} [Neighbors of sub-query]
The neighbors of a sub-query $s$ consist of $s$ itself, its parent sub-queries, and its child sub-queries: 
\begin{align} 
\mathrm{neighbors}(s) = \mathrm{parent}(s) \cup \mathrm{children}(s) \end{align} 
\noindent
where $\mathrm{parent}(s)$ denotes the set of sub-queries whose root nodes are parents of the root node of sub-query $s$,
while $\mathrm{children}(s)$ denotes the set of sub-queries whose root nodes are children of the root node of sub-query $s$.
\end{definition}
\noindent
In Figure~\ref{fig:subq_over_query}, parents and children of $s_3$ are indicated by solid-lined rectangles. The parents are sub-queries indicated by rectangles that cover red rectangles in $q_2$ and $q_3$, and the children are sub-queries indicated by sub-queries within the red rectangle, respectively. 

The neighbor-based sub-query exploration enumerates promising sub-tree candidates that balance the greater utility of materialization (achieved by enumerating parent sub-queries) and the higher potential for reuse across multiple queries (achieved by enumerating child sub-queries).

\subsubsection{Iterative Optimization}
\label{sssec:iteration}
Starting from the initial candidate sub-queries,
we explore the solution space by adding neighbor sub-queries and then choose new candidate sub-queries that maximize the objective function.
Let $C_t$ denote the candidate sub-queries at iteration $t$ ($t > 0$), and let $E_t$ denote the extended sub-queries obtained by adding neighbor sub-queries to $C_t$, defined as follows:
\begin{align} E_{t} = \bigcup_{s \in C_t} s \cup \mathrm{neighbors}(s) \end{align}
\noindent
where $C_0$ is the initial solutions selected by the methods described in Section~\ref{sec:initial}. 
Then, the new candidate sub-queries $C_{t+1}$ are computed by solving an ILP over the extended sub-queries $E_{t}$:

\begin{align} C_{t+1} = \mathrm{ILP}(\bmQ, E_{t}) \end{align}
\noindent
The function $\mathrm{ILP}(\bmQ, E_{t})$ selects the solutions to the view selection problem when the whole candidate sub-query set $\bmS$ is localized to a subset $E_{t}$, which corresponds to $L(\bmQ, E_{t})$ in Definition~\ref{def:loss}.
This localization is a key advantage of the local search approach, as it significantly reduces the number of decision variables ($y_{ij}$ and $z_j$) compared to the whole sub-query set $S$, that is $|E_{t}|<<|\bmS|$.
The iteration continues until no further improvement can be achieved in the objective function.

\subsection{Incremental View Maintenance Cost} \label{ssec:view_maintenane}
This section describes the details of the incremental view maintenance cost $m_j$ for a materialized view of subquery $s_j$ in response to update operations.
Since the workload contains update operations to the base tables, the materialized views must be maintained to reflect these changes.
Numerous studies~\cite{incremental:RPAI,DBToaster,incremental:tempura,10.1145/170035.170066} have explored incremental view maintenance, and some of these techniques have even been integrated into commercial systems, such as Oracle, SQL Server, and Tempura~\cite{incremental:tempura}.
Therefore, we focus on computing the cost of incremental view maintenance rather than that of full refresh-based view maintenance.

\subsubsection{Notations}
The incremental view maintenance cost in response to updates on base tables is expressed using a cost function $\mathsf{cost_{IVM}}(\Delta)$, where the argument $\Delta$ denotes the delta view expression for given update operations on base tables. 
$\Delta$ is expressed as an algebraic expression of the view, in which each base table is updated with its corresponding delta table that reflects incremental update operations.
As an example of a single-table view, $\Delta$ is expressed using a delta table, such as $\Delta A$ for Table~$A$, and the incremental view maintenance cost is denoted as $\mathsf{cost_{IVM}}(\Delta A)$.
If a view is defined as a multi-table view, for example, for a given view $A \bowtie B$ and an incremental update operation $\Delta B$ to the base table $B$, the incremental view maintenance cost is denoted as $\mathsf{cost_{IVM}}(A \bowtie \Delta B)$.

\subsubsection{Cost Estimation for Incremental Join View Maintenance}
In order to solve the view selection problem, we need to estimate the view maintenance cost ($m_j$) in Definition \ref{def:loss}.
However, since the database stores only the base tables and does not materialize the views in advance, the view maintenance cost must be estimated without physically generating the views.
We begin by decomposing the incremental view maintenance into three constituent operations, and then describe how to estimate their costs.

\paragraph{Decomposing incremental view maintenance cost}
First, the incremental view maintenance cost is calculated as the sum of the costs of three operations\footnote{If the update operations on the base table are insertion, the maintenance cost consists only of 1) and 3), because we can skip 2).}:
1) computing changes to the view,
2) identifying records to update in the view, and
3) applying the changes to the view.
Since the view maintenance cost of a single-table view is intuitive, we describe the incremental view maintenance cost by focusing specifically on binary join views. This approach can be easily extended to views defined by multi-table joins, as they can be represented as sequences of binary join operations in the execution plan.

Consider a binary join view, $V = A \bowtie B$, and an incremental delete operation $-\Delta B$ on Table $B$.
We omit the cases for incremental insert and update operations, as they can be straightforwardly derived from the delete case.
Let $\Delta V = A \bowtie \Delta B$. The incremental view maintenance cost is decomposed into the costs of the three operations and is expressed as follows.
\begin{align}
    \mathsf{cost_{IVM}}(A \bowtie \Delta B)
    = \mathsf{cost}(A\bowtie \Delta B) 
    + \mathsf{cost}(\sigma_{\Delta V}(V))
    + \mathsf{cost}(V -\Delta V)\label{eq:ivm_cost}
\end{align}

\paragraph{Estimating incremental view maintenance cost:}
Then, we estimate the cost of the above three operations as follows.

\smallskip
\noindent
{\bf{1) Computing changes to the view:}}
$\mathsf{cost}(A\bowtie \Delta B)$ is approximated as 
the cost of querying $A \bowtie B$ times $\frac{\Delta B}{|B|}$ (where $|B|$ is the number of records in B), 
because we only need to compute ratio $ \frac{\Delta B}{|B|}$ of $A \bowtie B$.
\begin{align}
    \mathsf{cost}(A \bowtie \Delta B) \simeq \mathsf{cost_{read}}(A \bowtie B) \times \frac{\Delta B}{|B|}
\end{align}

\smallskip
\noindent
{\bf{2) Identifying records to update in the view:}}
The cost of identifying $\Delta V$ in the materialized view $V$ is approximated as 
the cost of evaluating $\sigma_{\Delta B}(B)$ times
$\mathsf{fanout_{A \bowtie B}}(B\rightarrow V)$,
where $\mathsf{fanout}_{A \bowtie B}(B \rightarrow V)$ denotes the fanout (i.e., output cardinality) from $B$ to $V$ when $V$ is obtained by $A \bowtie B$.
\begin{align}
 \mathsf{cost}(\sigma_{\Delta V}(V)) \simeq \mathsf{cost_{read}}(\sigma_{\Delta B}(B)) \times \mathsf{fanout}_{A \bowtie B}(B \rightarrow V)
\end{align}
For example, if tables $A$ and $B$ are in a one-to-many relationship (i.e., one $A$ record corresponds to multiple $B$ records) and the view is defined using natural join, then $\mathsf{fanout}_{A \bowtie B}(B \rightarrow V) \simeq 1$.  
Conversely, if they are in a many-to-one relationship, then $\mathsf{fanout}_{A \bowtie B}(B \rightarrow V)\simeq |A|/|B|$.
If the view is defined using ad-hoc join, we use a simple histogram-based approach commonly used in relational databases.
It is our future work to employ machine-learning-based cardinality techniques~\cite{NeuroCard, FactorJoin} for more precise fanout estimation.

\smallskip
\noindent
{\bf{3) Applying the changes to the view:}}
The cost of applying the changes ($-\Delta V$) to $V$, $\mathsf{cost}(V -\Delta V)$, is approximated as 
the cost of deleting $\Delta B$ from the base table $B$ times 
$\mathsf{fanout}_{A \bowtie B}(B\rightarrow V)$.

\begin{align}
    \mathsf{cost}(V -\Delta V)\simeq \mathsf{cost_{write}}(-\Delta B) \times \mathsf{fanout_{A \bowtie B}}(B\rightarrow V)
\end{align}
where $\mathsf{cost_{write}}_{B}(-\Delta B)$ denotes the cost of deleting $\Delta B$ from its base table. 

\subsection{Algorithm Description}
\label{ssec: algorithm}
We define $\bm{U}$ (sub-query utilities), $\bm{b}$ (sub-query footprints), $\bm{m}$ (sub-query maintenance costs), and $\bm{z}$ (sub-query for materialization) as vectors consisting of elements $U_j$, $b_j$, $m_j$, and $z_j$, respectively, for sub-query $s_j$ ($1 \leq j \leq |\bm{S}|$). 
We also define $\mathbf{X} \in \{0,1\}^{|\bm{S}| \times |\bm{S}|}$ as a sub-query containment matrix consisting of elements $x_{jk}$, which indicates whether $s_j$ contains $s_k$, and
$\mathbf{Y} \in \{0,1\}^{|\bm{Q}| \times |\bm{S}|}$ as a matrix consisting of elements $y_{ij}$, which indicates whether query $q_i$ uses the materialized view of sub-query $s_j$.

The view selection algorithm is outlined in Algorithm~\ref{algo:proposed}.
\begin{algorithm}[t!]
\caption{View selection optimization}
\label{algo:proposed}
\begin{flushleft}
\textbf{Input:} $\bm{Q}$ (queries), $\bm{S}$ (sub-queries), $\bm{U}$, $\bm{b}$, $\bm{m}$, $X$, and $B_{\text{max}}$ (storage budget)\\
\textbf{Output:} $\bm{z}$, $Y$
\end{flushleft}
\begin{algorithmic}[1]
\STATE $\bm{z} = \textbf{initial\_solution}_{B_{\text{max}}}(\bm{S}, \bm{U},\bm{b})$ \hfill \textcolor{black}{// initial solution selection}
\STATE $U_{previous} = 0 $

\WHILE{$true$}
    \STATE $C = \textbf{candidate\_views}(\bm{z})$ \hfill \textcolor{black}{// $\bm{z}$ to $C$ transformation}
    \STATE $E = \bigcup_{s \in C} s\cup \textbf{neighbors}(s)$ \hfill \textcolor{black}{// sub-query exploration}
    
    \STATE $\bm{z}, Y = \textbf{ILP}_{ B_{max}, X,\bm{m}, \bm{U}}(\bm{Q}, E)$ \hfill \textcolor{black}{// applying ILP}

    \STATE $U_{current} = \sum_{q_i\in \bm{Q}} \sum_{s_j\in \bm{S}} u_{ij} \cdot y_{ij} - \sum_{s_j\in \bm{S}} m_j \cdot z_j$

    \STATE \hfill\textcolor{black}{// iteration termination check}
    \IF{$U_{previous} >= U_{current}$} 
        \STATE \textbf{break}
    \ENDIF
    \STATE \hfill\textcolor{black}{// preparing for the next iteration}
    \STATE $U_{previous} = U_{current}$

\ENDWHILE
\STATE \textbf{return} $\bm{z}$, $Y$
\end{algorithmic}
\end{algorithm}
Initially, an initial candidate sub-queries (expressed using $\bm{z}$) is generated using one of the metrics topk-$F$, topk-$U$, or topk-$E$, subject to the storage budget (Line 1-2) as detailed in Section~\ref{sec:initial}.
Then, we conduct a local search and ILP until no further improvement can be achieved (Lines 3-14).
Line 5 corresponds to neighbor-based sub-query exploration (detailed in Section~\ref{sssec:neighbor}),
and Line 6 involves applying Integer Linear Programming (ILP) optimization for the current candidate sub-queries ($E$).
ILP updates the decision variables ($\bm{z}, Y$) that maximize the objective function in Definition~\ref{def:loss}. 
Lines 9-11 determine termination based on convergence criteria, and Line 15 outputs the final decision variables ($\bm{z}$, $Y$). 


\section{View Materialization and Query Rewriting}
\label{sec:mv_create_query_rewrite}

Once the optimization of view selection is completed, we generate the materialized views and rewrite the original queries using the views based on the decision variables ($\bm{z}, Y$) where $y_{ij}$ indicates whether query $q_i$ uses the materialized view $s_j$, and $z_j$ indicates whether $s_j$ is selected for materialization.

\smallskip
\noindent
{\bf View generation}.
To define the materialized view, we extract all referenced tables, join conditions, and selection conditions from the execution plan of each sub-query $s_j$ and construct a \texttt{CREATE MATERIALIZED VIEW} statement. 
\begin{itemize} 
\item \textbf{CREATE MATERIALIZED VIEW as [view_name]}
\item \textbf{SELECT clause}: All columns from the referenced tables are included. The duplicate column names between different tables are resolved via renaming.
\item \textbf{FROM clause}: All tables from the referenced tables. 
\item \textbf{WHERE clause}: All conditions from the referenced tables.
\end{itemize}
When materialized views are defined, they are subsequently materialized within the database.

\smallskip
\noindent
{\bf Query rewrite using views}.
Each query is rewritten as follows based on $y_{ij}$: query $q_i$ uses the materialized view $s_j$:
\begin{itemize} 
\item \textbf{SELECT clause}: Replace the columns of the base tables in $q_i$ with the corresponding columns in $s_j$. 
\item \textbf{FROM clause}: Replace the corresponding base tables in $q_i$ with $s_j$. 
\item \textbf{WHERE clause}: Remove the join and filter conditions in $q_i$ if they are included in $s_j$. 
\end{itemize}

\section{Experiments}
\label{sec:exp}

The purpose of the experiments is to compare the performance of the proposed method with those of the baselines, a naive ILP-based method (Naive ILP) and BIGSUBS \cite{bigsubs}.
In detail, we conducted four experiments based on the following objectives:

\begin{description}
\item [\textbf{Workload performance improvement}:] We evaluate the workload performance of the proposed method compared to the baselines.

\item [\textbf{Ablation study: Effectiveness of local search}:]
To evaluate the effectiveness of local search, we compare the performance of the proposed method with and without local search applied to three different initial solutions.

\item [\textbf{Ablation study: Effectiveness of initial solution selections}:]
\kaina{We compare the workload performance of the three initial solution selections: topk-$F$, topk-$U$, and topk-$E$}

\item [\textbf{Parameter sensitivity: Impact of insert operations}:]
We examine how the number of insert operations affects the optimization cost for the proposed method and the baselines. 

\item [\textbf{Parameter sensitivity: Impact of storage budget}:]
We analyze how the storage budget size affects the total view utility and the optimization time by comparing the proposed method and the baselines.

\end{description}

\begin{table}[t!]
	\centering
        \caption{Experimental environment}
	\label{table_environment}
	\begin{tabular}{|l|l|} \hline
		Machine & MacBook Pro (16-inch, 2024) \\ \hline
            OS & macOS Sequoia 15.4.1(24E263)\\ \hline
		CPU & \makecell[l]{Apple M4 Pro \\ (14-core CPU, 20-core GPU)}\\ \hline
		Memory & 48GB \\ \hline
		ILP solver & Gurobi 12.0 \\ \hline
            Database system & PostgreSQL 17.4 \\ \hline
		Database & IMDb \\ \hline
		Database size & 7085MB \\ \hline

	\end{tabular}
\end{table}

\subsection{Setup} 

\subsubsection{Benchmark} 
\label{benchmark}
We conduct the experiments using the Redbench\footnote{https://github.com/utndatasystems/redbench}, which is a recently published benchmark extended from the Join Ordering Benchmark (JOB)~\cite{job} 
by introducing query repetition. 
The benchmark runs on the IMDb database, consisting of 21 tables.
The query repetition is generated based on the real-world workload from Redset\footnote{https://github.com/amazon-science/redset}, a dataset of query metadata published by Amazon Redshift, so Redbench is suitable for workload evaluations. 

Redbench contains 1073 JOB-based queries. 
Among these, there are 83 unique JOB queries\footnote{While the complete JOB workload comprises 113 queries, Redbench selects a subset of 83 queries that involve between 6 and 11 joined tables to construct its benchmark workload. }, 
corresponding to distinct query templates. Those queries are categorized into ten repetition buckets based on query repetitiveness, which is defined as the proportion of queries that appeared in the query history.
In addition, we manually added synthetic insert operations to the benchmark by randomly selecting base tables, because the Redbench workload does not support update operations.



\begin{table}[t!]
\centering
\small
\caption{
Amortized total time (sec) between proposed method and baselines, including overhead across methods for different query repetition buckets indicating repetition ratio. The storage budget ($B_{\text{max}}$) is set to 50 MB. 
The bold indicates the lowest total time in each query repetition bucket, and the underline indicates the second lowest.
}
\label{table:ex_11}
\resizebox{\columnwidth}{!}{%
\begin{tabular}{@{}l r r r r@{}}
\toprule
\begin{tabular}{@{}c@{}}\textbf{Query repetition} \\ \textbf{buckets}\end{tabular} & 
\multicolumn{1}{c}{\textbf{w/o MV}} & 
\multicolumn{1}{c}{\textbf{Naive ILP}} & 
\multicolumn{1}{c}{\textbf{BIGSUBS}} & 
\multicolumn{1}{c}{\textbf{Proposed (topk-$F$)}} \\
\cmidrule(lr){2-2} \cmidrule(lr){3-3} \cmidrule(lr){4-4} \cmidrule(lr){5-5}
\hspace{0.5em}0\%--10\% & 55.75 & 25.55 & \underline{20.67} & \textbf{14.96} \\
10\%--20\% & 21.43 & \underline{3.85} & 7.55 & \textbf{2.00} \\
20\%--30\% & 89.60 & \underline{41.02} & 51.64 & \textbf{22.47} \\
30\%--40\% & 71.19 & 18.56 & \underline{17.18} & \textbf{12.78} \\
40\%--50\% & 144.91 & \underline{57.29} & 75.68 & \textbf{51.17} \\
50\%--60\% & 4.32 & 1.45 & \underline{0.60} & \textbf{0.60} \\
60\%--70\% & 85.12 & \underline{49.79} & \textbf{47.36} & 66.17 \\
70\%--80\% & 107.16 & \underline{32.39} & 53.83 & \textbf{30.99} \\
80\%--90\% & \textbf{70.19} & 94.83 & 90.21 & \underline{77.22} \\
90\%--100\% & 226.08 & \underline{115.94} & 143.46 & \textbf{92.62} \\
\midrule
\textbf{Total Time (s)} & 875.75 & \underline{440.68} & 508.18 & \textbf{370.98} \\
\bottomrule
\end{tabular}
}
\end{table}

\subsubsection{Metrics}
As the evaluation metrics, we use 1) the total time consisting of optimization time for materialized view selection, view materialization time, and workload execution time, and 4) total view utility defined in Equation~(\ref{U_j}). 
We also report the total time for each of the ten repetition buckets by distributing the total optimization time and view materialization time across the buckets in proportion to each bucket’s workload execution time relative to the total workload execution time.
Regarding the view utility, we use the EXPLAIN command in PostgreSQL for estimating the sub-query execution cost ($cost_{read}(\cdot)$) in the view utility and the update operation cost ($cost_{write}(\cdot)$).

\subsubsection{Environment}
Table~\ref{table_environment} describes the environment we used in the experiments. 
In operating PostgreSQL, we have adopted the default configuration settings.
Regarding the convergence criteria for the optimization process, we apply the same approach to all methods: we repeat the iteration until no further improvement can be achieved in the objective function.

\subsection{Experimental Results}

\subsubsection{Workload Performance Improvement}
\label{exp:workload_improve}

\kaina{
This experiment demonstrates that our proposed method (topk-$F$) provides the most effective and practical solutions for workload acceleration, delivering near-optimal query performance with minimal optimization overhead.

We evaluate the overall performance of our method in terms of total time and compare it with the baselines.
We use the topk-$F$ initial solution selection as the representative configuration for our proposed method.}
Assuming a practical usage scenario, we first applied the view selection optimization using the original 113 JOB queries as the training queries, which serve as the query templates for the Redbench workload. We then evaluated the workload performance using the 1073 JOB-based queries as the test queries that constitute the Redbench workload.
We generate 1000 synthetic insert operations. We set the default size of the storage budget to 50 MB, which is approximately 10\% of the total size required to materialize all queries of the original JOB benchmark.

\kaina{First, Table~\ref{table:ex_11} shows the total time results for each query repetition bucket.
The results indicate that our proposed method is not only fast overall, but also consistently effective across diverse query repetition buckets. In the lower-repetition buckets (0\%–50\%), Proposed (topk-$F$) demonstrates a marked performance advantage over the baselines. This suggests that the local search is effective in narrowing the search space: by prioritizing frequently appearing sub-queries, it explores efficiently promising regions of the solution space, achieving high-quality solutions without exhaustive exploration.
}

\kaina{Second, Table~\ref{table:ex_12_improved} presents the breakdown of the total time, which includes optimization time, view materialization time, and workload execution time.
Our proposed method (topk-F) maintains a high view utility (3272K), which is comparable to the near-optimal utility achieved by Naive ILP (3374K), while keeping the optimization overhead remarkably low (1.29 seconds). 
In contrast, Naive ILP produces high-quality views that yield the fastest workload execution (166.93 seconds), but its prohibitively high optimization time (79.75 seconds) limits its practicality for real-time or large-scale scenarios. BIGSUBS, on the other hand, incurs a negligible optimization time (0.73 seconds) but selects views with significantly lower utility (1855K), leading to a longer workload execution time (431.45 seconds).}

\begin{table}[t!]
  \centering
  \small
  \caption{Comparison of total workload rerun (Total time) for view selection, including optimization time for view selection, materialized view (MV) generation time, and workload execution time, together with the utility of selected views. All times are reported in seconds..}
  \label{table:ex_12_improved}
  \begin{tabular}{@{}l r r r r r@{}}
    \toprule
    \textbf{Method} 
    & \thead{Total Time \\ (s)} 
    & \thead{Opt. \\ Time (s)} 
    & \thead{MV Gen. \\ Time (s)} 
    & \thead{Workload \\ Exec. time (s)} 
    & \thead{MV \\ Utility (K)} \\
    \midrule
    w/o MV                 & 875.75 & 0     & 0   & 875.75   & 0    \\
    Naive ILP              & 440.68 & 79.75 & 194 & 166.93 & 3374 \\
    BIGSUBS                & 508.18 & 0.73  & 76  & 431.45 & 1855 \\
    Prop. (topk-$F$)       & \textbf{370.98} & 1.29  & 179 & 190.71 & 3272 \\
    \bottomrule
  \end{tabular}
\end{table}

\subsubsection{Ablation study: Effectiveness of Local Search}

This ablation experiment evaluates the effectiveness of the local search in the proposed methods. We compared the optimization time and the total utility obtained by the proposed methods with and without the local search. That is, the initial solutions are used as the selected views without using the local search. 
The number of insert queries was set to 1000. 

Table~\ref{table:ex3_50M} shows that local search provides a substantial benefit in increasing total utility for topk-$F$ and topk-$E$, with only a marginal additional time.
As an example of topk-$F$, local search improves total utility by a factor of 2.1, incurring only 1.18 seconds of additional optimization time: it is significantly smaller than the whole workload execution time of approximately 100 seconds (see Table~\ref{table:ex_11}).
Note, in the case of topk-$U$, the total utility remains unchanged, indicating that the initial solutions are as good as or better than the ILP-refined solutions, as evidenced by the fact that the iteration terminates after the first round.

\begin{table}[tb!]
    \centering
    \caption{
    Comparison of total view utility and optimization time with and without local search ($B_{\text{max}}$: 50MB). 
    $\mathit{diff}$    indicates the difference between the proposed method with and without local search (proposed - w/o local search).
    }
    \label{table:ex3_50M}
\begin{tabular}{lrr}
\toprule
\textbf{Method} & \textbf{Utility (K)} & \textbf{Optimization} \\
                &                      & \textbf{Time (sec)} \\
\midrule
proposed (topk-$F$)               & 3270 & 1.29 \\
w/o local search (topk-$F$)       & 1570 & 0.11 \\
$\mathit{diff}$                             & +1700 & +1.18 \\
\hline
proposed (topk-$U$)               & 3340 & 0.76 \\
w/o local search (topk-$U$)       & 3340 & 0.28 \\
$\mathit{diff}$                            & +0 & +0.48 \\
\hline
proposed (topk-$E$)               & 2710 & 0.75 \\
w/o local search (topk-$E$)       & 2280 & 0.02 \\
$\mathit{diff}$                            & +430 & +0.73 \\
\bottomrule
\end{tabular}
\end{table}

\subsubsection{Ablation study: Effectiveness of initial solution selections}
\label{ex_prop-comparison}
\begin{table}[t!]
\centering
\small
\caption{
Amortized total time (sec) for proposed methods, including overhead across all query repetition buckets indicating repetition ratio. The storage budget ($B_{\text{max}}$) is set to 50 MB. 
The bold indicates the lowest total time in each query repetition bucket, and the underline indicates the second lowest.
}
\label{table:ex_32}
\begin{tabular}{@{}l r r r@{}}
\toprule
\multirow{2}{*}{\begin{tabular}{@{}c@{}}\textbf{Query repetition} \\ \textbf{buckets}\end{tabular}} & \multicolumn{3}{c}{\textbf{Proposed}} \\
\cmidrule(lr){2-4}
& \multicolumn{1}{c}{\textbf{topk-$F$}} & \multicolumn{1}{c}{\textbf{topk-$U$}} & \multicolumn{1}{c}{\textbf{topk-$E$}} \\
\midrule 
\hspace{0.5em}0\%--10\% & \textbf{14.96} & \underline{16.79} & 26.78 \\
10\%--20\% & \textbf{2.00} & \underline{9.59} & 14.66 \\
20\%--30\% & \textbf{22.47} & 64.46 & \underline{63.42} \\
30\%--40\% & \underline{12.78} & \textbf{7.29} & 21.88 \\
40\%--50\% & \underline{51.17} & \textbf{28.06} & 69.98 \\
50\%--60\% & \underline{0.60} & \textbf{0.39} & 4.93 \\
60\%--70\% & 66.17 & \textbf{38.40} & \underline{41.76} \\
70\%--80\% & \underline{30.99} & \textbf{20.43} & 68.87 \\
80\%--90\% & 77.22 & \textbf{41.54} & \underline{59.09} \\
90\%--100\% & \textbf{92.62} & 261.94 & \underline{230.21} \\
\midrule
\textbf{TOTAL} & \textbf{370.98} & \underline{488.90} & 601.89 \\
\bottomrule
\end{tabular}
\end{table}
\begin{table}[t!]
  \centering
  \small
  \caption{Comparison of total workload rerun (Total time) for view selection, including optimization time for view selection, materialized view (MV) generation time, and workload execution time, together with the utility of selected views. All times are reported in seconds.}
  \label{table:ex_3}
  \begin{tabular}{@{}l r r r r r@{}}
    \toprule
    \textbf{Method} 
    & \thead{Total Time \\ (s)} 
    & \thead{Opt. \\ Time (s)} 
    & \thead{MV Gen. \\ Time (s)} 
    & \thead{Workload \\ Exec. time (s)} 
    & \thead{MV \\ Utility (K)} \\
    \midrule
    Prop. (topk-$F$)    & \textbf{370.98} & 1.29  & 179 & 190.71 & 3272    \\
    Prop. (topk-$U$)    & 488.90 & 0.76  & 110 & 497.59 & 3344 \\
    Prop. (topk-$E$)    & 601.59 & 0.75  & 113 & 387.89 & 2707 \\
    \bottomrule
  \end{tabular}
\end{table}

\kaina{
Having established the overall superiority of our proposed method, this experiment delves into a comparative analysis of the three distinct initial solution selections: topk-$F$, topk-$U$, and topk-$E$. The objective is to understand the trade-offs inherent in each heuristic and to justify the selection of topk-$F$ as our representative method. The experimental setup remains consistent with that in Section~\ref{exp:workload_improve}.

First, Table~\ref{table:ex_32} shows the total time results for each query repetition
bucket.
This reveals the distinct strengths of each selection. topk-$F$ excels in both the low-repetition (0\%-20\%) and high-repetition (90\%-100\%) buckets. This behavior is intuitive: a frequency-based heuristic is adept at identifying sub-queries that are either broadly shared across many distinct queries or are part of extremely common queries. In contrast, topk-$U$ demonstrates its strength in the mid-range repetition buckets (30\%-80\%), where it often outperforms topk-$F$. This suggests that topk-$U$ is particularly effective at identifying high-cost, high-impact sub-queries that may not be the most frequent but offer significant optimization potential.

Second, Table~\ref{table:ex_12_improved} provides the breakdown of the total time. 
The results indicate that topk-$F$ achieves the best total time, with a total time of 370.98 seconds. Although topk-$U$ identifies a set of views with the highest estimated utility (3344 K), this does not necessarily lead to faster workload execution. Instead, topk-$F$ yields the lowest workload execution time (190.71 s), demonstrating that selecting initial views based on their frequency of appearance results in a more practically effective set of materialized views for this workload. The slightly longer optimization time for topk-$F$ (1.29s) suggests its initial solution was further from the final local optimum, requiring more iterative refinement, yet the superior quality of the final view set justified this marginal overhead. The topk-$E$ selection, which aims to balance utility and storage, proves least effective, resulting in both the lowest utility (2707K) and the longest execution time. This suggests that normalizing utility by storage cost may lead the local search to converge on a suboptimal solution.

In conclusion, while each initial solution selection has its merits, topk-$F$ emerges as the most robust and well-rounded approach. Its ability to achieve the best total time by effectively reducing the actual workload execution time makes it the most practical choice for general-purpose workload acceleration.
}

\subsubsection{Parameter sensitivity: Impact of the number of insert operations}

\kaina{
This experiment evaluates how the number of insert operations affects the optimization cost of our proposed method and the baselines.}
We manually generated synthetic insert operations and changed the number of inserted records from 0 to 2,000 in increments of 100.

\begin{figure}[t!] 
    \centering 
    \includegraphics[width=\hsize]{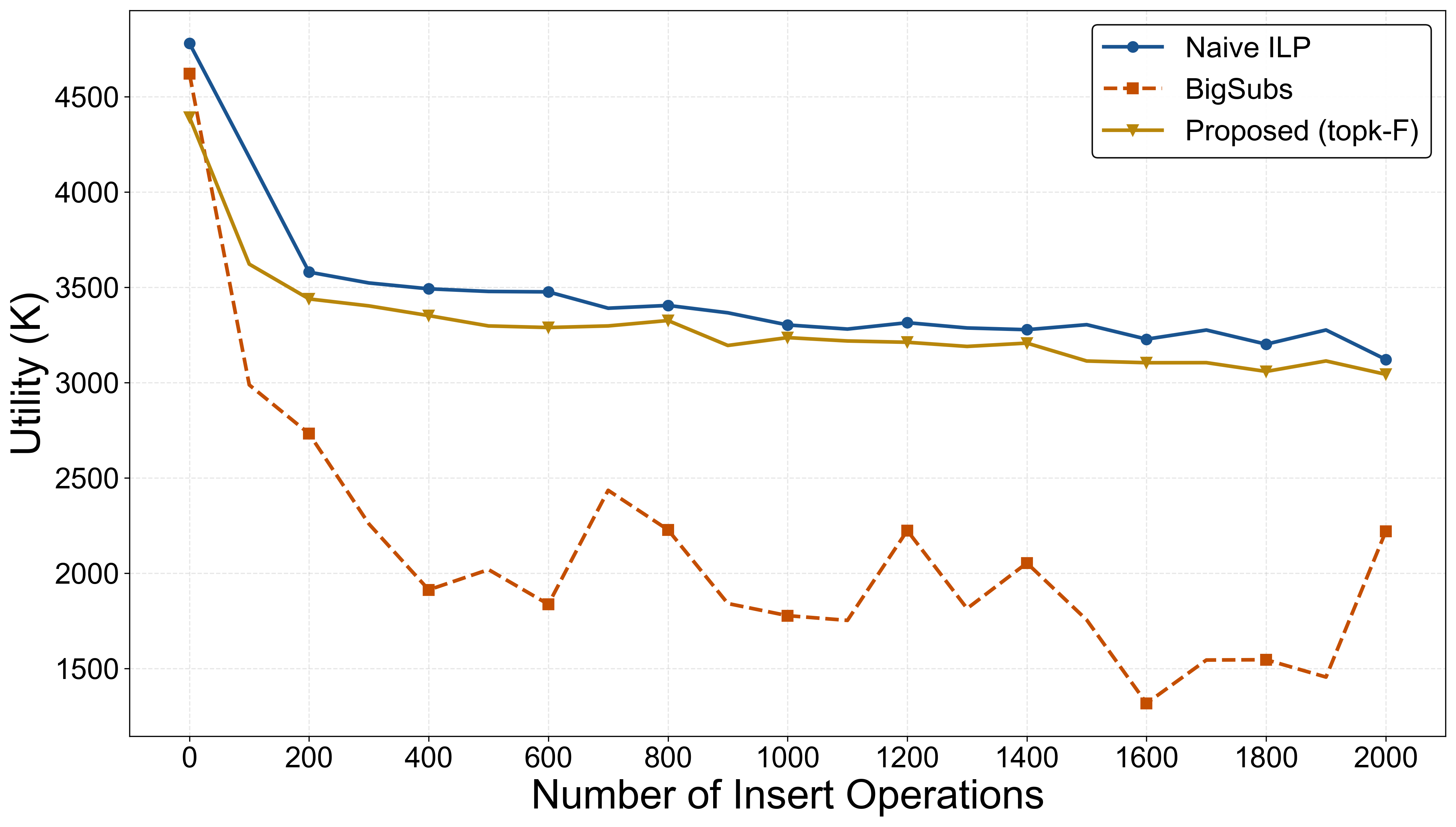} 
    \caption{Total view utility with respect to the number of insert operations. }
    \label{fig:ex_21} 
\end{figure}

\kaina{Figure~\ref{fig:ex_21} illustrates the total utility using the selected views as a function of the number of insert operations for our proposed method (topk-$F$), Naive ILP, and BIGSUBS. The utility of both topk-$F$ and Naive ILP gradually decreases as the number of inserts increases. This decline is expected, since the objective function explicitly penalizes view maintenance cost; as more insert operations occur, the cumulative maintenance overhead reduces the achievable utility. Nevertheless, topk-$F$ successfully adapts to this change by selecting views whose expected benefits remain close to optimal even under higher maintenance costs.

In contrast, the total utility achieved by BIGSUBS is generally the lowest and it does not exhibit a consistent decline but rather fluctuates. This instability in the total utility is attributed to its probabilistic approach, as noted in prior work~\cite{bigsubs}.
}

\subsubsection{Parameter sensitivity: Impact of storage budget}

This experiment analyzes how the storage budget size ($B_{\max}$) affects the total view utility and optimization time by comparing the proposed method with the baselines.
The number of insert operations was fixed at 1000, and the storage budget size $B_{max}$ was changed across 5 MB, 50 MB, and 500 MB.

\begin{figure}[h] 
    \centering 
    
    \begin{subfigure}[b]{0.48\textwidth} 
        \centering
        \includegraphics[width=\linewidth]{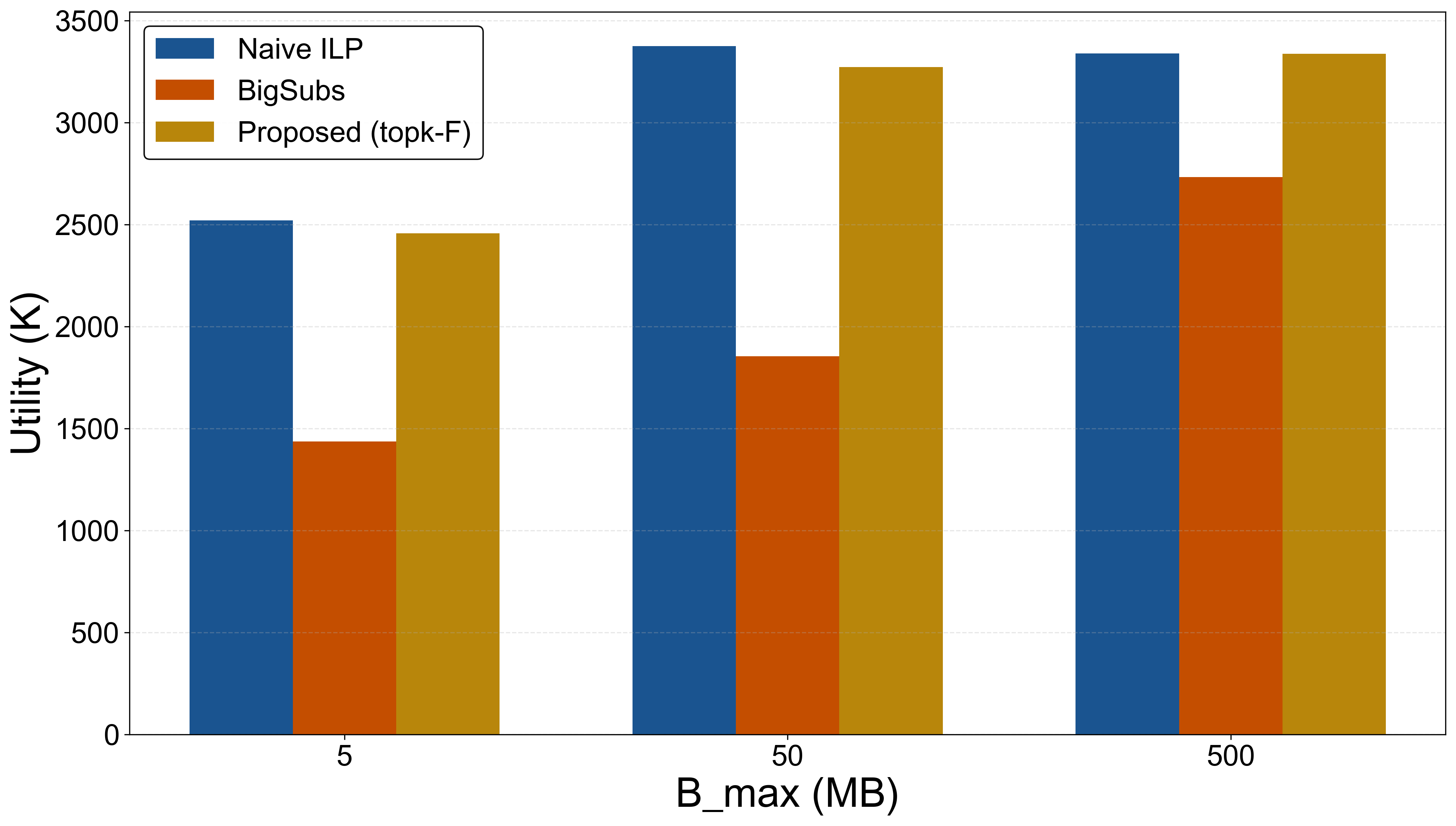}
        \caption{View utility}
        \label{fig:ex_22_u_sub}
    \end{subfigure}
    \hfill 
    \begin{subfigure}[b]{0.48\textwidth} 
        \centering
        \includegraphics[width=\linewidth]{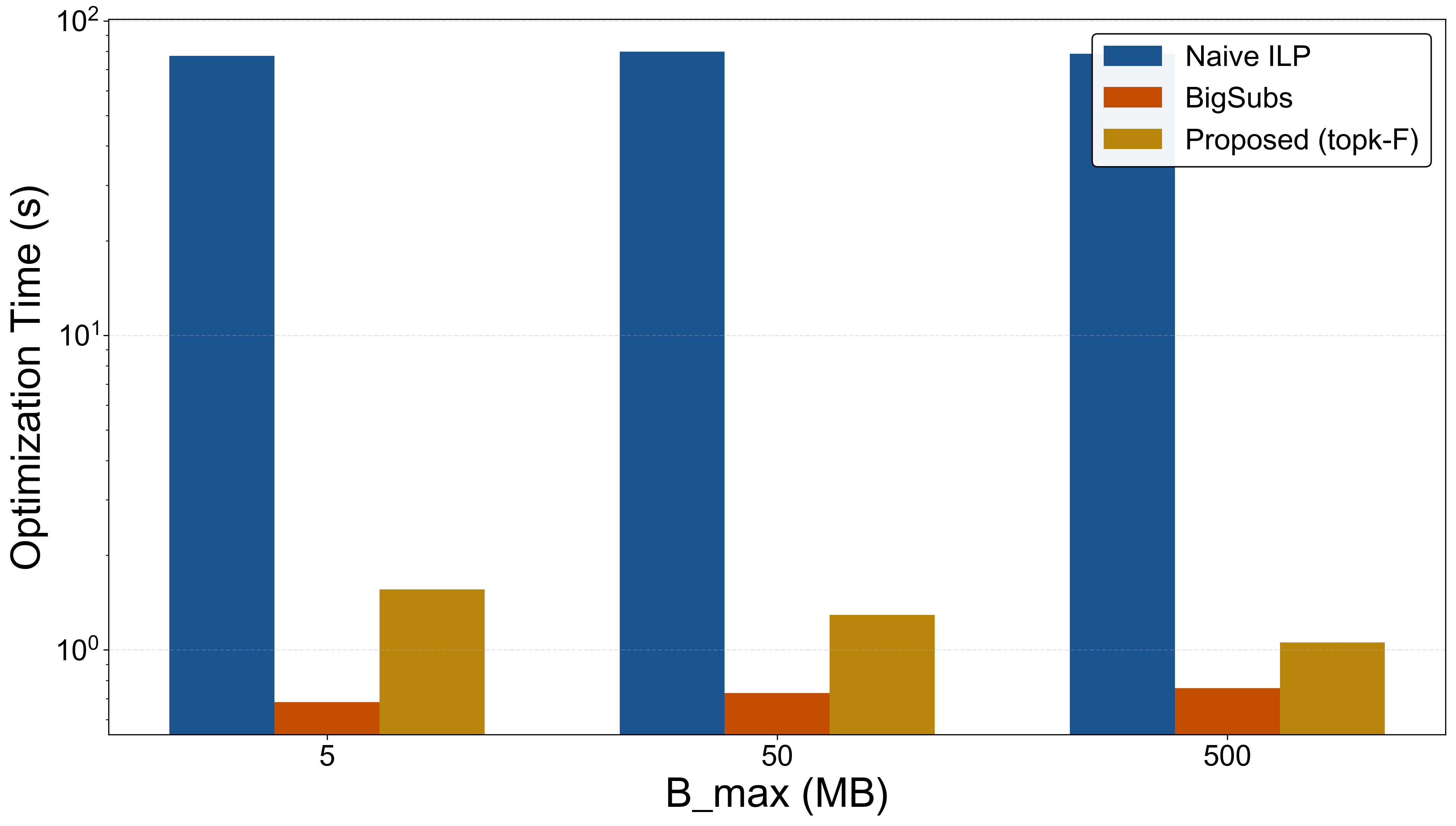}
        \caption{Optimization time}
        \label{fig:ex_22_t_sub}
    \end{subfigure}

    \caption{Comparison of total view utility and optimization time with respect to $B_{max}$ (5 MB, 50 MB, and 500 MB).} 
    \label{fig:ex_22_combined} 
\end{figure}

\kaina{Figure~\ref{fig:ex_22_u_sub} illustrates the total utility of the materialized views selected by each method. Our proposed method topk-$F$ consistently identifies the views with a utility nearly comparable to that of the Naive ILP approach, which provides a theoretical optimum. This stands in contrast to BIGSUBS, which selects a substantially less beneficial set of views across all budget constraints, indicating its lower selection accuracy. As expected, the utility for all methods increases with a larger $B_{max}$, as a greater budget allows for the materialization of more effective views.

Regarding optimization overhead, Figure~\ref{fig:ex_22_t_sub} shows the optimization time on a logarithmic scale. Both our proposed method and BIGSUBS achieve optimization times that are orders of magnitude faster than the Naive ILP approach. Although the optimization time of our method is almost the same as that of BIGSUBS, it achieves better overall performance while completing the optimization process within a few seconds. This minimal time investment is a stark contrast to the prohibitive computational cost of the Naive ILP method.

Overall, these results confirm that the proposed method effectively balances optimization quality and computational cost, delivering near-optimal performance under realistic resource constraints.}



\subsubsection{Validation of the Utility Metric}

\kaina{Furthermore, we confirmed the validity of our utility metric as a surrogate objective. For view sets selected by our proposed method, we observed a strong inverse correlation (Pearson's r = -0.920) between their total estimated utility and the resulting workload execution time. This validation underpins the effectiveness of our optimization approach.

}
\section{Related Work}
\label{sec:rela}

The selection of materialized views has been extensively studied, with comprehensive surveys~\cite{rela:MV_survey, halevy2001answering} providing overviews of a variety of strategies and their trade-offs, including view selection, query rewriting, and answering queries using views. 
A recent survey~\cite{rela:ML_survey} reviews the techniques for both AI4DB and DB4AI, including database configuration tuning, optimizer, and index/view advisor.
The techniques for view selection are categorized into heuristic-based approaches~\cite{rela:MV_cube, harinarayan1996implementing, gupta1997andor, theodoratos1997configuration, gupta2005view, rela:MV_heuristic} or optimization-based approaches that formulate the problem using integer linear programming (ILP)~\cite{papadomanolakis2007ilp, camacho2016pigreuse, bigsubs, wide-deep}.

\subsection{Heuristic-based approaches}
Classical techniques for view selection in data warehouses~\cite{gupta1997andor, gupta2005view, theodoratos1997configuration} clarify the theoretical aspect of view selection and propose greedy approaches.
These techniques rely on heuristics to search the space of candidate views and are effective for small workloads, but they suffer from limited scalability due to their local search nature and reliance on simplifying assumptions.

\subsection{ILP-based approaches}
ILP-based approaches have emerged as a principled strategy for optimizing view selection under complex constraints. These methods formulate the problem as an integer linear program where the objective is to maximize overall benefit or minimize total cost subject to storage and maintenance budgets.
Papadomanolakis and Ailamaki~\cite{papadomanolakis2007ilp} proposed one of the first ILP-based frameworks for database physical design, originally targeting index selection but extensible to materialized view selection. PigReuse~\cite{camacho2016pigreuse} applies ILP-based selection over AND/OR DAGs for Pig Latin scripts, enabling the reuse of common subplans. A prominent ILP-based system, BIGSUBS~\cite{bigsubs}, reformulates the view selection problem as a labeling task over a bipartite graph. It decomposes the large ILP into smaller subproblems and iteratively updates the set of materialized sub-queries. While scalable to large workloads, BIGSUBS relies on probabilistic heuristics and does not guarantee convergence to a global optimum.
Wide-deep~\cite{wide-deep} extends BIGSUBS by introducing a neural model to estimate view utility and uses deep reinforcement learning to guide view selection. However, this approach introduces additional complexity and may lead to instability in training and convergence.
In contrast, our work is the first work that employs local search, and it achieves high fast optimization speed and high quality of selected views. 

\subsection{Multi-query optimization}
Multi-query optimization (MQO) is a related but distinct problem that focuses on identifying opportunities to share computation across a batch of concurrent queries. MQO techniques typically use temporary views, which are materialized only during the execution of the batch.
Roy et al.~\cite{roy2000efficient} extend the Volcano optimizer with AND/OR DAGs to identify and reuse common subplans across queries, achieving significant cost savings.
Koch and Schlegel~\cite{koch2004algebraic} propose an algebraic approach that merges multiple queries into a unified representation, allowing efficient factoring of shared computations beyond pairwise reuse.
More recently, Anneser et al.~\cite{anneser2023autosteer} introduce AutoSteer, a learning-based framework that tunes optimizer parameters to improve performance across batches of similar queries, achieving up to 40\% speedup on engines like PostgreSQL and SparkSQL.
\section{Conclusion} \label{sec:conc}
We proposed a new method that incorporates incremental view maintenance cost directly into the optimization objective and applies local
search to efficiently explore the solution space. 
In order to apply
local search to the view selection problem, we define neighboring
solutions using sub-query containment, and select initial solutions
based on sub-query frequency, utility, or utility per storage unit.
We confirmed the practical effectiveness of our view selection method using a recent benchmark, RedBench. 

Future work includes the following directions: \begin{enumerate} \item Evaluating the effectiveness of materialized views on more generalized workloads \item Utilizing PostgreSQL’s incremental view maintenance functionality\footnote{https://wiki.postgresql.org/wiki/Incremental_View_Maintenance} \item Applying the wvlet framework for extracting logical plans, generating view definitions, and rewriting queries\footnote{https://github.com/wvlet/wvlet} \end{enumerate}

Since the JOB workload used in the experiments does not include common query patterns such as nested queries, it is necessary to investigate how the effectiveness of the proposed method and BIGSUBS changes when applied to more generalized workloads. Moreover, by implementing PostgreSQL's incremental view maintenance feature, we plan to evaluate the update cost during workload execution. In addition, since wvlet, a flow-based query language, allows direct extraction of logical plans, we aim to explore the feasibility of integrating our method with this framework.
\section{GenAI Usage Disclosure}

We gratefully acknowledge the use of generative AI tools in the preparation of this manuscript. Specifically, we used [ChatGPT-4o] for the following purposes:
\begin{itemize}
    \item Translation Assistance: The AI tool was employed for English proofreading and for translating our native language text into English. The authors thoroughly reviewed all translated content and made revisions to ensure accuracy and clarity.
    \item LaTeX Code Generation for Tables: Based on data collected and organized by the authors, the AI tool was used to generate LaTeX code for formatting tables. The authors verified the correctness of the generated code and ensured appropriate data presentation.
    \item Programming Support: During the development of the programs used in the experiments, the AI tool assisted in debugging and suggesting modifications. All code modifications proposed by the AI were reviewed, tested, and validated by the authors.
\end{itemize}

The authors take full responsibility for the final content and integrity of this research.
\bibliographystyle{ACM-Reference-Format}
\bibliography{sample-base}










\end{document}